\documentclass[letterpaper]{article}
\usepackage[utf8]{inputenc}
\usepackage{amsmath}
\usepackage{tabularx}
\usepackage[hidelinks]{hyperref}
\usepackage{graphicx}
\usepackage{pgfplots}
\usepgfplotslibrary{groupplots}
\pgfplotsset{width=\textwidth*0.5,compat=1.9}

\usepackage[
type={CC},
modifier={by-sa},
version={4.0},
]{doclicense}

\usepackage[sorting=none]{biblatex}

\title{Packet Compressed Sensing Imaging (PCSI): Robust Image Transmission over Noisy Channels
\footnotetext{\doclicenseThis Presented at ARRL/TAPR DCC 2020}}
\author{Scott Howard, Grant Barthelmes, Cara Ravasio,\\ Lisa Huang, Benjamin Poag, \& Varun Mannam}
\date{Department of Electrical Engineering, University of Notre Dame}

\begin{document}

\maketitle

\begin{abstract}
Packet Compressed Sensing Imaging (PCSI) is digital unconnected image transmission method resilient to packet loss. The goal is to develop a robust image transmission method that is computationally trivial to transmit (e.g., compatible with low-power 8-bit microcontrollers) and well suited for weak signal environments where packets are likely to be lost. In other image transmission techniques, noise and packet loss leads to parts of the image being distorted or missing. In PCSI, every packet contains random pixel information from the entire image, and each additional packet received (in any order) simply enhances image quality. Satisfactory SSTV resolution (320x240 pixel) images can be received in $\approx$1-2 minutes when transmitted at 1200 baud AFSK, which is on par with analog SSTV transmission time. Image transmission and reception can occur simultaneously on a computer, and multiple images can be received from multiple stations simultaneously - allowing for the creation of ``image nets." This paper presents a simple computer application for Windows, Mac, and Linux that implements PCSI transmission and reception on any KISS compatible hardware or software modem on any band and digital mode.
\end{abstract}

\tableofcontents

\section{Introduction}
Packet compressed sensing imaging (PCSI) is a solution to the technical challenge of transmitting a complete image to multiple receivers over a channel where each receiving station may miss different parts of the transmission due to channel noise. PCSI is implemented in a way such that a low-power microcontroller (e.g., Arduino) is capable of transmitting the image from challenging and  remote environments (e.g., high altitude balloon).

PCSI achieves these capabilities by using a technique known as `compressed sensing imaging," a computational method that allows one to reconstruct a complete image when only given a random selection of pixels from that image. Therefore, even after receiving only a single packet of random pixels, the receiver can begin to reconstruct the complete original image. Each additional packet received further increasing image quality. PCSI is, therefore, robust against packet loss. 

\begin{figure}
    \centering
    \includegraphics[width=0.9\textwidth]{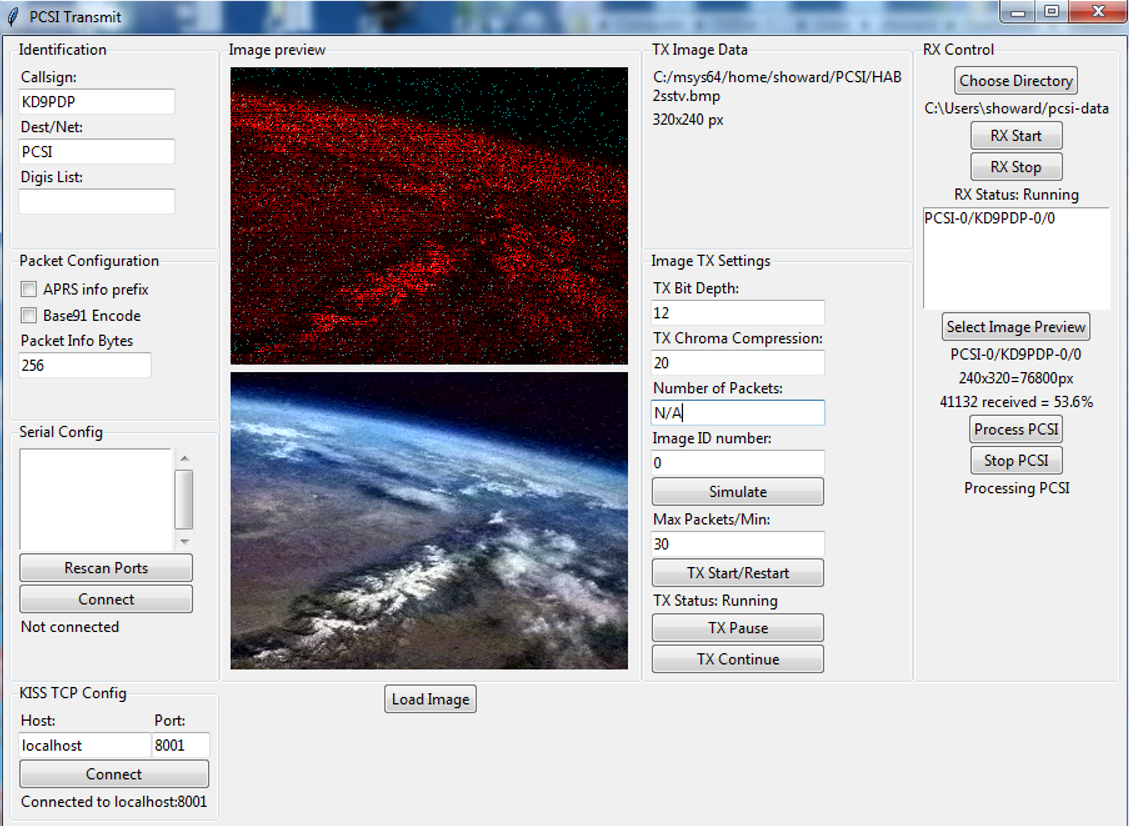}
    \caption{Demonstration of PCSI using pcsiGUI, an open-source tool written in Python for Windows, macOS, and Linux computers. Only 53.6\% of the packets were received (indicated by the red pixels in the top frame), yet the entire image can be reconstructed (bottom image). The software tool connects to any KISS compatible TNC or software modem and is capable of simultaneous (duplex) receiving/transmitting, and capable of receiving multiple images from multiple stations simultaneously.}
    \label{fig:screenshotl}
\end{figure}

The initial version of an open-source software tool \cite{pcsiwebsite} that implements sending and receiving PCSI in Python for Windows, macOS, and Linux is illustrated in Figure \ref{fig:screenshotl}. The image experienced $\approx 50\%$ packet loss, yet the image was fully reconstructed (bottom frame). In other techniques, 1/2 of the image would be completely missing.

\subsection{Radio Image Transmission Modes Comparison}
When transmitting an image, the operator is presented with many possible methods and must select an image transfer method that best suits the application. The following is a quick summary of popular techniques. 

Analog transmission can either encode an image's color values using frequency modulation (as in SSTV) or amplitude modulation (as in radiofax). In all cases, each point is transmitted sequentially; therefore, channel noise and signal loss leads to color distortion and pixel loss.

Digital techniques divide an image into individual pixels and transmit the quantized (i.e., represented in binary) values of each pixel's color channels as groups of data in packets. Each packet contains additional information that allows the receiver to check if the received packet was distorted during transmission (e.g., a ``checksum" as used in AX.25), and more recently, can correct for some errors using forward error correcting (FEC) codes (as used in FX.25). In both cases, unrecoverable errors lead to complete loss of pixel information for parts of the image.

Digital techniques also allow for the use of compression to send image data efficiently in fewer bits. A great example of this is slow scan digital video (SSDV) \cite{ssdvwebsite}. SSDV was developed by Philip Heron with the UK's high altitude ballooning community to transmit high-quality images. The technique takes images, converts them to a specific type of JPEG file, and transmits sections of the JPEG file in packets using FEC. This method transmits high quality images successfully; however, JPEG encoding and FEC generation is prohibitively complex for low-memory microcontrollers, and signal loss leads to missing parts of the image.

Hence, there is a need to develop a computationally simple image transfer method that is robust to signal loss and channel noise.

\subsection{PCSI's Predecessors}
In developing PCSI, we found that the idea of transmitting limited information over a noise channel while `filling in the blanks" at the receiving end were present in at least two other technologies.

\textbf{APRS Vision System}: At the 1997 DCC, Bob Bruninga (WB4APR) proposed the ``APRS Vision System" \cite{arpsvision}. That approach was an attempt to relay imaging information over the APRS network. The idea was to transmit APRS packets that contain increasingly higher amounts of spatial information content. The first packet contained an extremely low quality image, and each subsequent packet doubled the resolution. When the receiver feels the image is ``clear enough" (i.e., the receiver can "fill in the gaps" in the image), the receiver can tell the sender to stop. This way the receiver can possibly control or react to the image before the whole image became ``clear." However, this system required that every previous packet had been received perfectly for subsequent packets to be used, and the receiver needed to communicate with transmitter to indicate missed packets and to stop. Any missed packet caused the whole system to fail from that point forward.

\textbf{Hellschreiber}: This mode, developed in the mid 20th century, is a fascinating approach that enabled robust transmission of text over noisy channels \cite{hells}. Each character of text to be transmitted is converted in to a 7x7 pixel image of that character. Each character is then transmitted using extremely narrow bandwidth transmission (e.g., on-off CW keying). The receiver takes the received data and reconstructs the image. The operator then ``reads" the resulting image text. Fundamentally, the operator's trained character recognition neural network (i.e., their brain) does pattern recognition to ``fill the the gaps" in the image that was caused by channel noise and signal loss. The entire message can therefore be received even if some pixels were not received or if noise distorted some pixels.

\section{The Magic in the Math}
While it may seem ``too good to be true" to be able to fully reconstruct an entire image using only a few randomly selected pixel values, PCSI is made possible via the magic in the math. PCSI accomplishes this using two techniques: compressed sensing imaging and chroma compression.

\subsection{Compressed Sensing Imaging}
The mathematical concept of compressed sensing imaging exploits the fact that data can often be represented in ``sparse" domains. By ``sparse," we mean that ``most of the values are zero." For example, compare a photograph of a blooming garden during the day to a photograph of the night sky. The image of the garden will have lots of red, green, and blue color, and therefore, many non-zero values for the R, G, and B channels in the image. The night sky is mostly empty, with a few stars, planets, and the moon. The nigh sky image, therefore, has many pixels that have zero value in the R, G, or B channels. We would say the garden image is not sparse while the night sky is sparse.

The magic happens when you convert the image from the spatial domain (i.e., a photograph) to another ``domain." For example, you can perform a mathematical operation on an image called the discrete cosine transform (DCT). The resulting DCT ``image" ends up containing all the same information from the original image, just arranged in a different way. You can then do an inverse discrete cosine transform (IDCT) on the ``DCT image," and you will restore the original image accurately. Why would we want to take the DCT of an image? We do this because practically all images are ``sparse" after taking the DCT.\footnote{Technically, the image is sparse in the DCT domain because cameras typically massively ``oversampling" an image.} In other words, most of the values of the DCT of an image are zero! Both the garden and night sky look sparse after taking the DCT!

How can we use the fact that most images are sparse after taking the DCT? This is the magic of PCSI. You ask the computer to find the DCT of an image such that:
\begin{enumerate}
    \item The values for the DCT it finds has a lot of zeros in it (i.e., is sparse).
    \item When you reconstruct the image using the values for the DCT the computer found, the resulting real image closely matches whatever values of the pixels that were received
\end{enumerate}

Mathematically, you are asking the computer to do something similar to ``basis pursuit denoising" to find the simplest image (i.e., the fewest non-zero DCT values) that also matches the pixels that you have received so far. You can do that by using a computer program to find the values of the DCT of the final image ($\mathbf{X}$) such that it gives you the smallest value of the following expression:

\begin{equation}
    \sum_n{ | \text{IDCT}(\mathbf{X})_n- b_n |^2} + C\sum{|\mathbf{X}|}
\end{equation}
where $b_n$ is the value of the $n$th pixel that was received, and $C$ is a scaling factor (typically in the range of 3-5). The first term is the sum of the squared error between the values of the pixels in the reconstructed image and the value of the pixels that were actually received. Ideally, this will be zero. The second term is the L1 norm, which is adding up all the values of $\mathbf{X}$, and minimizing that term is a good method for finding a sparse $\mathbf{X}$. Once you find $\mathbf{X}$, you can take the IDCT of $\mathbf{X}$ to find the reconstructed image.

While the current PCSI reference implementation does use the above basis pursuit technique, it is just one of many ways to reconstruct an image from a collection of random pixels. PCSI actually does not require any particular reconstruction algorithm as there may be variations in the method that yield superior results.

\subsection{Chroma Compression and Color Depth}
PCSI image transmission speed is increased by reducing the bit depth of an image and by utilizing chroma compression. These techniques are described below.
\begin{itemize}
    \item \textbf{Bit Depth:} Reducing an image from 24 bit color (8-bit in each of R, B, G) to 12 bit color (or any other color depth) is trivial, may be acceptable for many applications, and therefore is an option in PCSI.
    \item \textbf{Chroma compression:} The human eye has 20-times high density of rods (greyscale photoreceptors) than cones (color photoreceptors), and therefore detects greyscale with better spatial resolution than color. It is therefore not necessary to transmit the same resolution for both luma (brightness) and chroma (color information). JPEG exploits this by representing an image in the YCbCr color space and sub-sampling the chroma channels (Cb and Cr) relative to the luma channel (Y). PCSI uses the same general concept and achieves chroma compression by sending a combination of full-color (YCbCr) and greyscale-only (Y) pixels in each packet. This step leads to the receiver receiving more Y channel pixels than Cb and Cr channel pixels. Each channel, separately, undergoes the compressed sensing basis pursuit to reconstruct the original channel, and the channels are then converted back to RGB. The resulting image appears to have much higher quality for the same number of packets.
\end{itemize}

\section{Implementation: The PDP Specification}
PCSI requires packet payloads such that each individual payload contains all the information necessary to reconstruct a single image. To achieve that, the pseudo-random datagram payload (PDP) specification has been developed. Version 1.0.0 is described below.

\subsection{What is the PDP?}
PDP is a specification for the payloads of data packets such that each packet contains all the information needed to reconstruct a single image. An image is then transmitted as the collection of datagrams (i.e., packets in a connection-less network). Unlike other packetized image transmission formats, the pixels contained in a packet are selected in a pseudo-random, yet deterministic, way. This allows images to be restored using compressed sensing techniques regardless of packet loss. 

\subsection {PDP Specification Scope}
The PDP spec merely defines the packet payload for the transmission of a single image. It can be used in any packet protocol or digital mode. Framing is independent of the specification. This allows for the separation of a ``session" (consisting of a sending station sending one or more images) from the minimal content required for a single image. The ``session" information is in the framing; the image information is in the payload. The payload is designed to be similar to SSDV.

For example, a PDP can be placed as the payload in:
\begin{itemize}
    \item AX.25 amateur radio packets. Transmitted using any mode (e.g., AFSK, PSK, etc.) Therefore it is compatible with APRS, TNCs, digipeaters, software modems (direwolf, fldigi, soundmodem, etc.). Example implementation is in Section \ref{section:AX.25framing}.
    \item SSDV-style framing done in a KISS TNC compatible way. Example implementation is in Section \ref{section:SSDVframing}.
    \item UDP or TCP (although the benefits of PCSI provide more benefit to multicast UDP packets than connected TCP packets).
\end{itemize}

\subsubsection{AX.25 Framing}\label{section:AX.25framing}
While not part of the PDP spec, an example of using AX.25 UI framing \cite{ax25} of a PDP is given in Table \ref{table:AX.25framing}. This is easily compatible with existing TNCs. This framing adds at least 20 bytes of overhead. 

\begin{table}[tbp]
    \centering
\begin{tabularx}{\textwidth}{ |l|l|X| } 
 \hline
 \textbf{Name} & \textbf{Size (bits)} & \textbf{Description} \\ \hline\hline
 Flag & 8 & HDLC flag `$\mathtt{0x7E}$' \\ \hline
 Dest. Address & 56 & Callsign of intended receiver OR alias of an image net, encoded following AX.25 spec. `PCSI` recommended for general use. \\ \hline
 Source Address & 56 & Sender's callsign encoded following AX.25 spec. `PCSI' recommended for general use. \\ \hline
 Digi Addresses & $d\times56$ & $d$ optional digipeater addresses, encoded following AX.25 spec. \\ \hline
 Control & 16 & `$\mathtt{0x03F0}$' indicating UI frame with no response requested, and no layer 3 implemented  \\ \hline
 \textbf{PDP} & $N\times 8$ & \textbf{PDP data}, $N\leq 256$ \\ \hline
 FCS & 16 & CRC-CCITT\\ \hline
 Flag & 8 & HDLC flag `$\mathtt{0x7E}$' \\
 \hline
\end{tabularx}
\caption{Example AX.25 framing that could be used for PCSI.}
\label{table:AX.25framing}
\end{table}

\subsubsection{SSDV-style Framing}\label{section:SSDVframing}
While not part of the PDP spec, a simple session framing of PDP can be done in a way that is compatible with existing KISS hardware and software TNCs. An example is seen in Table \ref{table:SSDVframing}. This example framing is designed to be easy to use with any KISS TNC. One would simply send the concatenated ``Packet Identifier + Callsign + PDP" and let the TNC add the flags and do the checksum. This framing adds at least 9 bytes of overhead. 

\begin{table}[tbp]
    \centering
\begin{tabularx}{\textwidth}{ |l|l|l|X| } 
 \hline
 \textbf{Offset (bytes)} & \textbf{Name} & \textbf{Size (bytes)} & \textbf{Description} \\ \hline\hline
 0 & Flag & 1 & HDLC flag `$\mathtt{0x7E}$' \\ \hline
 1 & Packet Identifier & 1 & ASCII `v' = `$\mathtt{0x76}$' \\ \hline
 2 & Callsign & 4 & Base-40 encoded callsign following SSDV encoding convention \\ \hline
 6 & \textbf{PDP} & $N \leq 256$ & \textbf{PDP data} \\ \hline
 N+6 & Checksum & 2 & CRC-CCITT\\ \hline
 N+8 & Flag & 1 & HDLC flag `$\mathtt{0x7E}$' \\
 \hline
\end{tabularx}
\caption{Example SSTV-style framing that could be used for PCSI.}
\label{table:SSDVframing}
\end{table}

\subsubsection{Framing Comparison}
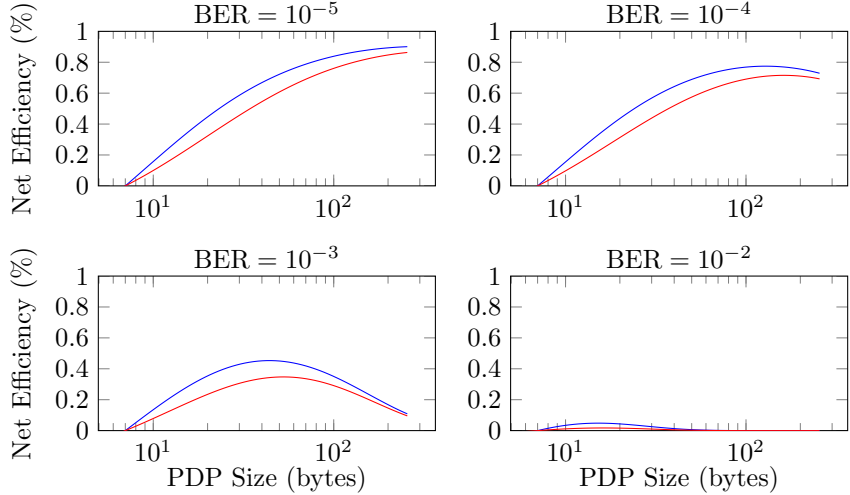
\begin{figure}[tbp]
\centering
\begin{tikzpicture}
\begin{groupplot}[
      group style={group size=2 by 2, vertical sep=1.2cm},
      width=\textwidth*0.5, height=\textwidth*0.3,
      xmode=log,
      ymin=0,
      ymax=1,
      xlabel shift = -5pt,
      title style={yshift=-1.5ex}
    ]
    \nextgroupplot[ylabel = Net Efficiency (\%),
    legend entries={SSDV-Style,AX.25},
    legend columns=-1,
    legend to name=named,
    title = {$\text{BER}=10^{-5}$}]

\addplot [
    domain=1:256, 
    samples=100, 
    color=blue,
    ]
{(1-10^(-5))^(x*8+9*8) * (x-7)/(9+x)};

\addplot [
    domain=1:256, 
    samples=100, 
    color=red,
    axis lines = left,
    xlabel = PDP Size (bytes),
    ylabel = Overhead (\%),
]
{(1-10^(-5))^(x*8+20*8) * (x-7)/(20+x)};

    \nextgroupplot[legend pos=south east, title = {$\text{BER}=10^{-4}$}]

\addplot [
    domain=1:256, 
    samples=100, 
    color=blue,
    ]
{(1-10^(-4))^(x*8+9*8) * (x-7)/(9+x)};
\addplot [
    domain=1:256, 
    samples=100, 
    color=red,
    axis lines = left,
    xlabel = PDP Size (bytes),
    ylabel = Overhead (\%),
]
{(1-10^(-4))^(x*8+20*8) * (x-7)/(20+x)};


    \nextgroupplot[ylabel = Net Efficiency (\%),xlabel = PDP Size (bytes), legend pos=north east, title = {$\text{BER}=10^{-3}$}]

\addplot [
    domain=1:256, 
    samples=100, 
    color=blue,
    ]
{(1-10^(-3))^(x*8+9*8) * (x-7)/(9+x)};

\addplot [
    domain=1:256, 
    samples=100, 
    color=red,
    axis lines = left,
    xlabel = PDP Size (bytes),
    ylabel = Overhead (\%),
]
{(1-10^(-3))^(x*8+20*8) * (x-7)/(20+x)};

    \nextgroupplot[xlabel = PDP Size (bytes),legend pos=north east, title = {$\text{BER}=10^{-2}$}]

\addplot [
    domain=1:256, 
    samples=100, 
    color=blue,
    ]
{(1-10^(-2))^(x*8+9*8) * (x-7)/(9+x)};

\addplot [
    domain=1:256, 
    samples=100, 
    color=red,
    axis lines = left,
    xlabel = PDP Size (bytes),
    ylabel = Overhead (\%),
]
{(1-10^(-2))^(x*8+18*8) * (x-7)/(18+x)};

\end{groupplot}
\end{tikzpicture}

\caption{Comparison of net bit efficiency versus PDP payload size for both SSDV-style and AX.25 framing.}\label{fig:EfficiencyVsPDP}
\end{figure}

Both AX.25 framing and SSDV-style framing can be used. AX.25 is more powerful as it can leverage existing packet radio infrastructure at the cost of of larger overhead. However, if channel bit error rate (BER) is high (as is common in longer-distance HF modes), smaller packets are more likely to be successfully received. The lower overhead SSDV-style framing may be superior in this case. This trade-off is explored in Figure \ref{fig:EfficiencyVsPDP}. The net efficiency (percent of each transmitted bit that will successfully transmit pixel-level image information to the receiver)  is calculated as the product of the probability that the entire packet will be received properly and the percentage of bits in a packet that correspond to pixel information.
\begin{align}
    \text{Net Efficiency AX.25} &= \frac{x-7}{x+20}\times(1-\text{BER})^{8x+160}\\
    \text{Net Efficiency SSDV-Style} &= \frac{x-7}{x+9}\times(1-\text{BER})^{8x+72}\\
\end{align}
where $x$ is the total PDP length in bytes and the term $x-7$ comes from the fact that the PDP has a 7 byte header as described in \ref{section:PDPspec}.

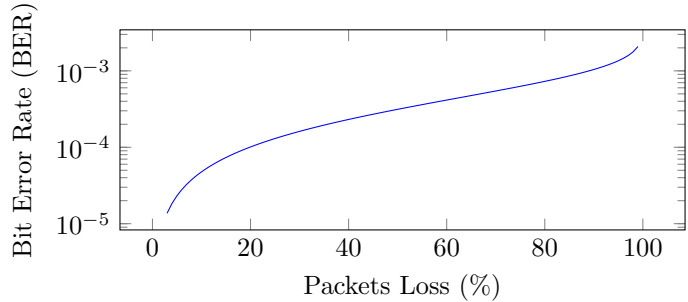
\begin{figure}[tbp]
\centering
\begin{tikzpicture}
\begin{axis}[
      width=\textwidth*0.75, height=\textwidth*0.35,
      ymode=log,
      ylabel = Bit Error Rate (BER),
      xlabel = Packets Loss (\%),
      ]

\addplot [
    domain=2:99, 
    samples=100, 
    color=blue,
    ]
table {BER.table};  

\end{axis}
\end{tikzpicture}
\caption{BER vs packet loss for AX.25 frames with 256 byte payloads.}\label{fig:BERvsLoss}
\end{figure}

Results from Figure \ref{fig:EfficiencyVsPDP} give guidelines for ideal PDP length.  First, find the approximate BER by estimating packet loss for an AX.25 packet with a 256 byte payload using the equation:
\begin{align}
    \text{BER}= 1-(1-L/100)^{1/2192}\label{eq:BER}
\end{align}
where $L$ is packet loss in percent. If the packet loss percentage is known (or can be estimated), BER can be found using Equation \ref{eq:BER}, which is depicted by Figure \ref{fig:BERvsLoss}.


Now based on the estimated BER, you can choose the appropriate framing style and PDP size (refering to Fig. \ref{fig:EfficiencyVsPDP}):
\begin{itemize}
    \item For low $\text{BER}\leq10^{-5}$ Environments: Framing style does not matter that much, and payloads should be the full 256 bytes long.
    \item For $\text{BER}\approx 10^{-4}$ Environments: Framing style does not matter that much, and payloads should be 130 bytes long.
    \item For $\text{BER}\approx 10^{-3}$ Environments: SSDV-Style framing increases efficiency (and speed) by $\approx 25\%$ compared to AX.25. Payloads should be 40-50 bytes long.
    \item For $\text{BER}\approx 10^{-2}$ Environments: AX.25 is practically unusable; SSDV framing will barely be usable. Payloads should be 10-11 bytes long.
\end{itemize}

\subsection{PDP Specification Details}\label{section:PDPspec}
Since each packet contains information of the whole frame, each packet \textbf{MUST} be the same size of every packet in an image (same number of pixels per packet). Total packet size is determined by the framing protocol used. For example, AX.25 packet payloads are 256 bytes by default. The payload contains the following data transmitted in order as described by Table \ref{table:PDPSpec}.

\begin{table}[tbp]
    \centering
\begin{tabularx}{\textwidth}{ |p{1.5cm}|p{1.5cm}|p{1.5cm}|X| } 
 \hline
 \textbf{Offset (bits)} & \textbf{Name} & \textbf{Size (bits)}  & \textbf{Description} \\ \hline\hline
 0 & Image ID & 8 & Identifies unique images within a PCSI session. (uint8)\\ \hline
 8 & Rows & 8 & Number of lines in the image divided by 16. (4096 lines max, uint8) \\ \hline
 16 & Columns & 8 & number of columns in the image divided by 16. (4096 columns max, uint8) \\ \hline
 24 & Packet ID & 16 & used as the starting point of the pseudorandom pixel list. (uint16) \\ \hline
 32 & Number of YCbCr Pixels & 8 & Number of full color pixels transmitted in this packet. (uint8)\\ \hline
 40 & Color depth & 8 & Color depth encoded as (color depth/3 -1). e.g., 24bit color = 7. This only uses 3 bits, so there are 5 bits available for future use. (uint8) \\ \hline
 48 & YCbCr Pixel Data & (Number of YCbCr Pixels) * (Color bit depth) & Full color (YCbCr) pixels listed first as a binary stream. For example, if color is transmitted as 12-bit color, each pixel is 12-bits long with the first 4 corresponding to the Y channel, the next four corresponding to the Cb channel, and the final 4 corresponding to the Cr channel. \\ \hline
 48 + (YCbCr Pixels) * (Color bit depth) & Y-only Pixel Data & N & Black and white (Y only) pixels follow in a binary stream of Y values encoded as a uint with the same bit depth as a single channel of the YCbCr pixels. \\ \hline
  & Zero padding & Z & Zero padding for byte alignment as needed. \\
 \hline
\end{tabularx}
\caption{PDP Specification Version 1.0.0}
\label{table:PDPSpec}
\end{table}

\subsubsection{Packet Payload Preparation}

Given a bit mapped image to transfer, follow the following procedures

\begin{enumerate}
    \item Using a pseudo-random number generator (see Section \ref{section:PRNG}), generate the sequence of pixels to be transmitted.
    \item Given the number of bits available in the payload (e.g., AX.25 UI frames have 256 bytes minus 7 bytes of image info equals 1992 bits total), the desired chroma compression level, and the desired color bit depth to transmit, determine the list of pixels to transmit that will be full color and solely back and white.
    \begin{enumerate}
        \item All packets consist of the same number of pixels (e.g., every packet for an image has exactly 25 YCbCr pixels and 75 Y only pixels for a total of 100 pixels. You can choose whatever numbers you want, as long as they are the same for every packet of the image).
    \end{enumerate}
    \item Prepare the packet payload
    \begin{enumerate}
        \item Convert full color pixels to YCbCr per ITU-T T.871 \cite{jpeg} and black and white only pixels to Y as per the same spec.
        \item If color bit depth is being reduced, approximate the value to be transmitted using rounding. For example, the 8 bit number 200 will be represented as the 4 bit number round(200/255*15)=12.
    \end{enumerate}
\end{enumerate}

\subsubsection{Pseudo-random Number Generation for Picking Pixels}\label{section:PRNG}
Compressed sensing imaging requires that the measurements are uncorrelated in the sparse domain that is used to reconstruct the image. Taking random samples ensures this condition, however, both the transmitter and receiver need to know which pixel values correspond to which pixels in the image.
To do this, PCSI uses a Linear Congruential Generator \cite{lcg} as a deterministic pseudo-random number generator using GCC's default coefficients ($\text{modulus}=2^{31}$, $a=1103515245$, $c=12345$, starting with a $\text{seed}=1$).
The pseudo-random number generator is then used with the modern Fisher Yates shuffle algorithm \cite{fyshuffle} to generate a random list of the pixels to be sent.
See reference code for details. This approach will allow all receivers and the transmitter to generate identical lists of order that pixels will be transmitted. Since every packet has the same number of pixels, the packet ID will give the receiver the starting pixel number from which the list of pixels received in the packet can be extracted.

Pixels are indexed column-first as seen in C, not row first as is typically done in Python. You therefore have to transpose a matrix before selecting and assigning pixels if you are working in Python.

\subsubsection{PCSI Payload base91 Encoding}
If you are transmitting over channels that only allow printable ASCII text, the entire PDP can be converted to base91 as described below. This is a combination of APRS base91 and basE91 \cite{base91}. Compared to basE91, the method used in PCSI is simpler and deterministic at the cost of slightly more overhead.

While there are 13 bits or more to convert, read in 13 bits
Convert those 13 bits to two ASCII bytes using [floor(bits/91)+33] for first and [bits\%91+33] for the second byte.
Next, if there are fewer than thirteen and more than seven bits available (the end of the stream), read in and zero pad (to the right, i.e., least significant bits) the remaining bits so that there are 13 bits total.
Convert those 13 bits to two ASCII bytes using [floor(bits/91)+33] for first and [bits\%91+33] for the second byte
If there are 6 or few bits remaining:
Read in and zero pad (to the right, i.e., least significant bits) the remaining bits so that there are 6 bits total.
Convert those 6 bits to one ASCII byte using bits+33.

\subsection{Reconstructing PCSI Images}
There is no specification or standard on how to reconstruct the images. Users can experiment with different methods and find what is appropriate. The reference implementation follows these steps \cite{rtpyrunner}:

\begin{enumerate}
    \item Decode all the pixel values and pixel numbers from as many packets as have been successfully received.
    \item For each color channel (Y, Cb, Cr), use OLW-QN for basis pursuit \cite{OWL} to find the discrete cosine transform (DCT) coefficients that best fit the received data and minimizes the L1 norm. This is the key to compressed sensing!
    \item After finding the DCT coefficients, use the inverse DCT to generate the color channels for the image.
    \item Convert from YCbCr to RGB, and save the image.
\end{enumerate}

\section{Future Work}
PCSI is available for use any band and KISS compatible TNC or software modem. Now that it has been demonstrated, some additional features can be explored:
\begin{itemize}
    \item\textbf{Low-power micro-controller transmission client:} PCSI transmission is computationally simple enough to be performed on low-memory micro-controllers. High-altitude balloons would benefit from integrating PCSI transmission with Arduino radios such as the HamShield (\url{https://inductivetwig.com/}).
    \item \textbf{PCSI aggregation server:} Since different stations can receive different packets, and increasing the packets increases image quality, a centralized server can be used to aggregate packets to improve image quality. This system would be similar to the Automatic Picture Relay Network (APRN \cite{aprn}) and to what is done with SSDV to receive images from high-altitude balloons that move out of range of the original receiving system \cite{ssdvaggregator}.
    \item \textbf{Integrate PCSI with APRS:} While transmitting entire PCSI images over APRS channels would severely strain the network, APRS could be leveraged to announce ongoing transmission or upcoming ``ImageNets" on non-APRS frequencies. APRS frequency objects can be transmitted following the conventions of the Automatic Frequency Reporting System (AFRS) \cite{afrs} and APRS Local Frequency Info Initiative \cite{localinfo}.
    
\end{itemize}

\section*{List of Terms}

\begin{tabularx}{\textwidth}{r X} 
\textbf{Notation} & \textbf{Meaning} \\
 AFSK & Audio Frequency Shift Keying  \\ 
 APRS & Automatic Packet Reporting System  \\ 
 ASCII & American Standard Code for Information Interchange \\
 AX.25 & Amateur X.25   \\ 
 BER & Bit Error Rate \\
 CRC & Cyclic Redundancy Check \\
 CW & Continuous Wave \\
 DCC & Digital Communications Conference \\
 DCT & Discrete Cosine Transform \\
 FCS & Frame Check Sequence \\
 FEC & Forward Error Correction \\
 FX.25 & Extension to AX.25 with FEC \\
 GCC & GNU Compiler Collection \\
 GUI & Graphical User Interface \\
 HDLC & High-Level Data Link Control \\
 HF & High Frequency \\
 IDCT & Inverse Discrete Cosine Transform \\
 ITU & International Telecommunication Union \\
 JPEG & Joint Photographic Experts Group \\
 KISS & Keep It Simple, [Silly] \\
 PCSI & Packet Compressed Sensing Imaging \\ 
 PDP & Pseudo-random Datagram Payload (PDP) \\
 OLW-QN & Orthant-Wise Limited-memory Quasi-Newton \\
 RGB & Red, Green, Blue \\
 SSDV & Slow Scan Digital Video \\
 SSTV & Slow Scan Television  \\
 TCP & Transmission Control Protocol \\
 TNC & Terminal Node Controller \\
 UDP & User Datagram Protocol \\
 UI & User Interface \\
 Y$\mathrm{C_B C_R}$ & Luma, blue-difference and red-difference chroma components \\
\end{tabularx}

\printbibliography[title={Bibliography}]

\end{document}